# Growth of epitaxially oriented Ag nanoislands on air-oxidized Si(111)-(7×7) surfaces: Influence of short range order on the substrate

Anupam Roy [a], K. Bhattacharjee [b, $], J. Ghatak [b, #] and B. N. Dev [a, *]

[a] Department of Materials Science, Indian Association for the Cultivation of Science, Jadavpur, Kolkata 700032, India

[b] Institute of Physics, Sachivalaya Marg, Bhubaneswar 751005, India

**Abstract**

Clean Si(111)-(7×7) surfaces, followed by air-exposure, have been investigated by reflection high energy electron diffraction (RHEED) and scanning tunneling microscopy (STM). Fourier transforms (FTs) of STM images show the presence of short range (7×7) order on the air-oxidized surface. Comparison with FTs of STM images from a clean Si(111)-(7×7) surface shows that only the $1/7^{th}$ order spots are present on the air-oxidized surface. The oxide layer is ~ 2-3 nm thick, as revealed by cross-sectional transmission electron microscopy (XTEM). Growth of Ag islands on these air-oxidized Si(111)-(7×7) surfaces has been investigated by *in-situ* RHEED and STM and *ex-situ* XTEM and scanning electron microscopy. Ag deposition at room temperature leads to the growth of randomly oriented Ag islands while preferred orientation evolves when Ag is deposited at higher substrate temperatures. For deposition at 550°C face centered cubic Ag nanoislands grow with a predominant epitaxial orientation $[1\bar{1}0]_{Ag} \parallel [1\bar{1}0]_{Si}$, $(111)_{Ag} \parallel (111)_{Si}$ along with its twin $[\bar{1}10]_{Ag} \parallel [1\bar{1}0]_{Si}$, $(111)_{Ag} \parallel (111)_{Si}$, as observed for epitaxial growth of Ag on Si(111) surfaces. The twins are thus rotated by a 180° rotation of the Ag unit cell about the Si [111] axis. It is intriguing that Ag nanoislands follow an epitaxial relationship with the Si(111) substrate in spite of the presence of a 2-3 nm thick oxide layer between Ag and Si. Apparently the short range order on the oxide surface influences the crystallographic orientation of the Ag nanoislands.

$ Present address: Max-Planck Institute for Microstructure Physics, Weinberg 2, D-06120 Halle, Germany.

# Present Address: Dept. of Material Science and Engineering, National Cheng Kung University, 1 University Road, Tainan, Taiwan

* Corresponding Author. Dept. of Materials Science. Indian Association for the Cultivation of Science. Jadavpur. Kolkata – 700 032. India. E-mail: msbnd@iacs.res.in. Phone: +91 33 2473 4971 (Ext. 200). Fax: +91 33 2473 2805

# 1. Introduction

Oxidation of clean silicon surfaces has been an important topic in surface science. For oxygen adsorption on clean Si(111)-(7×7) surfaces, early work has indicated that the (7×7) reconstruction fades away with increasing oxygen exposure and the oxidized layer is amorphous [1-3]. In other words, it was thought that adsorption of one or two monolayers of oxygen were enough to remove the reconstruction of the clean silicon surface. It was later demonstrated by reflection high energy electron diffraction (RHEED) studies that the (7×7) pattern persists even after high air exposure which results in oxide film growth of more than 1 nm in thickness. However, the authors have concluded that the (7×7) structure is not present at the oxide surface, but at the selvedge of the Si(111) substrate, i.e., below the oxide layer [4]. In our present work on oxidized Si(111)-(7×7) surfaces, obtained by air exposure, in addition to RHEED we use scanning tunneling microscopy (STM) which directly probes the surface and demonstrates that a short range (7×7) structure is present on the oxide surface. This (7×7) order is not apparent directly in the STM images. Only Fourier Transform (FT) of the images reveals this order. This is a short range order. Next, we investigate the aspect of growth of metallic nanoislands, e.g., Ag nanoislands, on this oxidized Si(111)-(7×7) surfaces with short range order, in order to explore whether this short range order on the oxide surface has any influence on the growth behaviour. Ag growth on a clean Si surface follows the layer-plus-island or Stranski-Krastanov growth mode [5-9]. However, on an oxidized surface, because of its much lower surface free energy, growth of Ag islands is expected. In that case, a short range order on the substrate surface would be expected to influence the growth of Ag islands. Here we explore the short-range order on air-oxidized Si(111)-(7×7) surfaces and the crystallographic orientation(s) of the Ag nanoislands grown on them.

Metal nanoparticles with size comparable to their electron mean free path have many unusual properties and functionalities. They serve as model systems for exploring quantum and classical coupling interactions and as building blocks of practical applications [10]. It is well known that the optical, electronic and biological properties of Ag nanoparticles are strongly dependent on their sizes and shapes [11,12]. In particular, the oriented and shaped Ag nanoparticles have drawn much attention because of their applications in bionanotechnology. Being the best conductor among metals Ag nanoparticles may facilitate more efficient electron transfer than other metal nanoparticles in biosensors. That is why Ag nanoparticles are the most intensively studied and applied metal nanoparticles in fabricating biosensors [13].

Growth of metallic islands on clean as well as oxidized Si surfaces is important for several applications. For example, Ag island films epitaxially grown on Si(111)-(7×7) surfaces as well as on surface-oxidized Si(111) substrates have shown enhanced infrared absorption for some organic materials adsorbed on them [14]. Ag nanoparticles on $SiO_2$ have been found to be a novel catalyst [15]. Single electron tunneling devices require two insulating tunneling barriers on two sides of a metallic dot; the oxide on which the metallic island is grown provides a tunneling barrier [16].

The metal nanoparticles themselves can be multiply twinned nanoparticles (MT-NPs) or single crystalline nanoparticles (SC-NPs). Chemical activity of MT-NPs show a substantial difference compared to SC-NPs. SC-NPs have many superior qualities regarding biological and chemical sensing, nanoscale mechanical characteristics, electron-phonon scattering etc. [10]. Because of lower surface and volume energies, for most of the noble-metal nanoparticles, during the nucleation and growth, it is very difficult to avoid twinning planar defects [17,18]. During the growth of a crystal if the crystal is kept under a stress or at an elevated temperature or pressure, two or more inter-grown crystals can be formed in a symmetrical manner - these symmetrical intergrowths of crystals are the twinned crystals. Twinning defects originate from the boundaries between multiple crystalline phases. Physicists have found that for sizes below 4 nm, Ag nanoparticles go through a structural transition [19]. As the constituent atomic arrangement changes, it affects most of the particle properties. The structure follows five-fold symmetric noncrystallographic structures of multiply twinned particles (MTPs) (decahedra in the 1-3 nm range and icosahedra for sizes below 1 nm). Metal MTPs (or MT-NPs) grow preferentially with [111] facets, five-fold symmetry and elastic deformation of the constituent tetrahedral domains [18]. Molecular dynamic simulation studies have predicted that Ag MTPs adopt the bulk, face-centered cubic (fcc) structure (SC-NPs) beyond a critical size of 8 nm [20]. It has been reported that most of the Ag nanoparticles formed on hydrogen terminated Si(001) surfaces are MT-NPs and they transformed to epitaxial fcc nanocrystals (SC-NPs) upon annealing [21] which has been further investigated by Sato et. al. [22]. Recently we have reported growth of fcc Ag nanoislands (SC-NPs) on oxidized surfaces of initially clean Si(001)-(2×1) substrates [23]. In spite of the presence of a 2-3 nm silicon oxide layer between the Ag nanoislands and the Si(001) substrate, the Ag islands were found to have an epitaxial orientation as if they have directly grown on a Si(001) substrate. However, what actually has influenced the epitaxial orientation of Ag nanoislands on the oxidized surface was not clear. In the present work we have investigated this aspect.

Here we have investigated the effect of a substrate surface (modified prior to the growth of nanoparticles), as well as the substrate temperature during deposition, on the structure of metal nanoparticles. Prior to Ag deposition, the substrate, Si(111)-(7×7) surface, has been modified, following Roy et. al. [23], by a method of interface modification by exposing clean Si(111)-(7×7) reconstructed surfaces to air, which gives rise to the formation of a thin $SiO_x$ layer on the clean Si surface. We have investigated the order on the oxidized Si(111) surfaces. Ag deposition onto this surface results in growth of nearly spherical, small and non-epitaxial fcc Ag nanoparticles. With rise in substrate temperature during Ag deposition, the evolution of preferentially oriented twinned fcc Ag particles from the randomly oriented Ag nanoparticles has also been investigated. Results of *in-situ* RHEED, STM and *ex-situ* scanning electron microscopy (SEM) and cross-sectional transmission electron microscopy (XTEM) investigations are presented.

## 2. Experimental details

Ag deposition was carried out under ultrahigh vacuum (UHV) condition in a custom-built molecular beam epitaxy (MBE) chamber where the base pressure inside the growth chamber was ~ $2 \times 10^{-8}$ Pa. A RHEED set-up is attached with the MBE system for in-situ monitoring of reconstruction and growth. A commercial UHV variable temperature scanning tunneling microscope (VTSTM, Omicron Nanotechnology, Germany) is connected to the growth chamber for in-situ characterizations of the samples. This system was described elsewhere [24]. P-doped n-Si(111) samples (oriented within ± 0.5°) with a resistivity of 5-20 Ohm-cm was loaded in the UHV chamber. Atomically clean, reconstructed Si(111)-(7×7) surfaces were prepared by usual heating and flashing procedure. The samples were first degassed at about 600°C for 12-14 hours followed by a flashing at ~ 1200°C for one minute. The substrate was then cooled down to room temperature (RT) and Si(111)-(7×7) surface reconstruction was observed by in-situ RHEED and STM. The samples were then exposed to air for 3-4 hours by bringing them out via the load lock chamber. This is expected to form a thin $SiO_x$ film on the Si(111)-(7×7) surface. Following this, the samples were again introduced into the UHV growth chamber and degassing of the sample was carried out at around 500°C for 2 hours. This procedure was carried out to remove adsorbed gases from the sample surface due to air-exposure without removing the thin oxide film formed on the Si(111)-(7×7) surface. After cooling down the samples to room temperature (RT), the surfaces were investigated by RHEED and STM. Different samples with Ag coverages upto 10 monolayers (ML) were prepared for different substrate temperatures during deposition (keeping all other growth parameters same) on $SiO_x$-covered Si(111)-(7×7) surfaces from a PBN (pyrolytic boron nitride) crucible. [Here a monolayer of Ag is defined to be $1.5×10^{15}$ atoms/cm$^2$ as used in our previous study [25] of Ag growth on clean Si(111) and in other studies [26]]. Substrate temperatures were varied from RT to 550°C. The growth rate of Ag was kept the same all over the experiment (2.1 ML/min) and the base pressure inside the UHV growth chamber during deposition was $8.5×10^{-8}$ Pa. The amount of Ag deposition during growth was measured by a quartz microbalance which has been pre-calibrated by Rutherford backscattering spectrometry (RBS) measurements [using a 2 MeV He$^+$ ion beam from the Pelletron accelerator]. The post-growth investigations of the samples were carried out by in-situ RHEED operated at 15 kV, STM and ex-situ SEM [with 5.0 kV electrons using a Hitachi S-2300 microscope] and high resolution XTEM [with 200 kV electrons using a JEOL-2010 (UHR) electron microscope] studies.

## 3. Results and Discussions

STM images [Fig. 1(a) and (b)] reveal a clean Si(111)-(7×7) surface. The clean surface has the presence of monatomic steps and terraces [Fig. 1(a)] and the surface has the (7×7) atomic arrangement which is the most stable surface reconstruction for Si(111) [Fig. 1(b)]. The step-terrace structure is preserved when the surface is exposed to air leading to the growth of a thin oxide layer on top of the surface [Fig. 1(c)]. STM images following 10 ML of Ag deposition on this air-oxidized Si(111) surface at room temperature (RT), reveal the formation of densely packed Ag islands. One such STM image is shown in Fig. 1(d).

Oxidation of passivated Si surfaces in air is a slow process related to desorption rate of the passivating species and somewhat dependent on crystallographic orientation of the surface [27]. Investigations regarding the oxidation process on H-terminated Si(001) surfaces show a 0.5 nm thick oxide layer growth on the surface at RT after keeping the sample for one day in air where the step-terrace structure was found to be preserved even after the oxidation [28]. However, clean surfaces, when exposed to air, undergoes the oxidation process comparatively rapidly (with a sticking coefficient of $10^{-4}$ - $10^{-1}$) [29]. In the present case a 3-4 hour exposure to air was sufficient to form a thin oxide layer on top of the surface. For oxygen adsorption, earlier it has been reported that the low energy electron diffraction pattern from the clean Si(111)-(7×7) surfaces fades away with increasing oxygen exposure and that the oxidized layer is amorphous [1-3]. Tabe *et. al.* [4], from their RHEED study have found that the (7×7) surface after air exposure (1 bar, 20h) still exhibits several (7×7) spots. The estimated thickness of the air-oxidized film corresponding to this exposure is 1.3 nm. They concluded that the (7×7) superstructure was not present at the oxide surface, but at the substrate selvedge buried under the oxide film. The fact that the energetic electron beam can penetrate the interface has apparently influenced that conclusion. In the present work, for the thin oxide layer (~ 2-3 nm as revealed from the cross-sectional transmission electron microscopy images presented later) grown by air-exposure, we use STM in addition to RHEED studies and observe that a short range order with the periodicity of the (7×7) reconstruction of the Si(111) surface is preserved on the oxide surface. This is discussed in the next paragraph.

Fig. 2 shows STM images of the Si(111)-(7×7) surface following air-exposure. The corresponding Fourier transforms (FTs) of the STM images are also shown in Fig. 2. Although the oxide layer grown by direct air-exposure is not likely to be as clean as the UHV-grown oxide layer, presence of the $1/7^{th}$ order spots indicates that the periodicity of the underlying Si(111)-(7×7) unit cell remains on the oxide surface grown by air-exposure, even though the oxide layer is ~ 2-3 nm thick. That the spots in the FTs in Fig. 2 are indeed the $1/7^{th}$ order fractional spots is demonstrated by comparing FTs from the clean Si(111)-(7×7) surface. Fig. 3 shows STM images from a clean Si(111)-(7×7) reconstructed surface and their corresponding FTs. Higher order spots are also seen in the FTs because of the presence of long range order on the clean Si(111)-(7×7) surface. Sizes of the STM images (scan area) in Fig. 3(a) and 3(b) are the same as those of Fig. 2(a) and 2(b) respectively. As in Fig. 2, the $1/7^{th}$ order spots appear closer to the center with decreasing scan area. The positions of the $1/7^{th}$ spots are exactly the same for the clean as well as for the oxidized surface obtained by air-exposure. Though the STM images from the oxidized surface do not show the order on the surface, FTs of the images reveal the presence of a short range order. For thermally grown silicon oxide layers inside a UHV chamber, Fujita *et. al.* also observed in the FTs of STM images that the periodicity of the Si(111)-(7×7) structure remains on the oxide surfaces [30].

A RHEED pattern of a clean reconstructed Si(111)-(7×7) surface is shown in Fig. 4(a) for $[1\ 1\ \bar{2}]_{Si}$ incidence. These surfaces were modified by exposing to air, and thus forming a thin oxide layer at the top. Fig. 4(b) shows a RHEED pattern from the same Si(111) surface after air-exposure followed by degassing. Although the clear (7×7) pattern has vanished, there is a weak remanence of some fractional order spots, consistent with the weak $1/7^{th}$ order spots seen in the FTs in Fig. 2. Moreover, the presence of some spots indicates the presence of some degree of order in the oxide layer rather than being amorphous. Ag was grown on these air-exposed surfaces. Growth was carried out at different substrate temperatures - from RT to 550°C – under UHV condition. Fig. 4(c) shows a RHEED pattern when 10 ML Ag was deposited at RT. The RHEED patterns show different features when Ag is grown at an elevated substrate temperature, e.g., at 350°C [Fig. 4(d)], at 400°C [Fig. 4(e)] and at 550°C [Fig. 4(f)]. We notice that well defined spot patterns have developed in Fig. 4(e) and 4(f). It is known that RHEED patterns from surface roughness or from microcrystalline/nanocrystalline three dimensional structures on surfaces are similar to diffraction patterns for transmission high energy electron diffraction (THEED). The conditions for the observation of THEED pattern in RHEED experiments are discussed in our previous work [31]. The RHEED patterns are explained below.

It is obvious that the crystalline quality of deposited Ag depends on the substrate temperature during deposition. After depositing 10 ML of Ag at RT, only broad Debye-Scherrer rings of Ag appear in the RHEED pattern while other previous spots are absent [Fig. 4(c)]. The ring diffraction pattern of Ag comes from the growth of randomly oriented Ag islands on the air-oxidized Si surfaces. The crystalline quality gradually improves with increasing substrate temperature. Spot-like Ag diffraction patterns start appearing when the substrate temperature during deposition increases. Fig. 4(d) shows the RHEED pattern from a sample when the temperature during deposition was kept at 350°C. In addition to the broad Debye rings, appearance of bright spots indicates the growth of textured polycrystalline 3D Ag islands. The diffraction pattern in Fig. 4(e) indicates the presence of an increasing number of preferentially oriented Ag islands. At 400°C growth temperature (Fig. 4(e)) Debye-Scherrer rings practically disappear and a set of bright and a set of weak spots appear. This indicates that the Ag islands have preferred orientation(s). At 550°C (Fig. 4(f)), the Debye-Scherrer rings disappear completely and a set of brighter spots become prominent in the RHEED pattern along with a set of weak spots. Considerable azimuthal alignment of the Ag islands is clear from the RHEED pattern. The set of bright RHEED spots is consistent with the epitaxial orientations – $[1\bar{1}0]_{Ag} \parallel [1\bar{1}0]_{Si}$, $(111)_{Ag} \parallel (111)_{Si}$ with an associated twin structure (to be discussed latter in more details along with simulation). This predominant parallel orientation was also observed for epitaxial growth of Ag(111) layers on clean Si(111) [32] and Ge(111) surfaces [33]. In the present case although there is an oxide layer present between the Ag islands and the Si surface, the epitaxial relationship indicates as if the oxide layer is not there. In other words, the short range order observed on the oxide surface apparently influences the crystallographic orientation of the Ag islands grown on them. For Ag growth on air-exposed Si(001)-(2×1) surfaces, Roy et. al. [23] has observed a similar epitaxial orientation of Ag islands: [1 1 0]$_{Ag}$ ∥ [1 1 0]$_{Si}$, (001)$_{Ag}$ ∥ (001)$_{Si}$.

Fig. 5 shows the RHEED images along the [1 $\bar{1}$ 0]$_{Si}$ azimuth from the same samples as in Fig. 4. RHEED pattern from a clean Si(111)-(7×7) surface is shown in Fig. 5(a). Fig. 5(b) shows the same surface oxidized upon air-exposure. The dependence of preferential orientation of Ag islands on the substrate temperature during deposition is shown in Fig. 5(c)-(f). The RT deposition shows random distribution of Ag islands and only broad Debye-Scherrer rings appear [Fig. 5(c)] as in Fig. 4(c). The bright spots start appearing when Ag is deposited at elevated temperatures. The Debye-Scherrer rings become weaker when the substrate temperature was increased to 350°C [Fig. 5(d), compare Fig. 4(d)]. The bright spots dominate over the Debye-Scherrer rings in the RHEED pattern when Ag is deposited at 400°C substrate temperature. At 550°C only spots from the deposited Ag film appear in the RHEED pattern [Fig. 5(f), compare Fig. 4(f)]. The RHEED pattern in Fig. 5(f), is indicative of Ag(111) islands with twinned structures. RHEED pattern simulation of a fcc(111) oriented crystal with twins for the [1 $\bar{1}$ 0] azimuth [34] resembles the pattern in Fig. 5(f). RHEED pattern for [1 1 $\bar{2}$] azimuth is not sensitive to this twin structure. It is noticed that the twinned structure evolves from 350°C onwards. Identification of the crystallographic orientation of the Ag islands on the basis of RHEED patterns for different azimuths is discussed in more details latter.

Fig. 6 shows how the crystalline Ag islands evolve with increase in the thickness of the Ag film when deposited directly on a hot substrate at a temperature of 550°C. The electron beam is incident along [1 1 $\bar{2}$]$_{Si}$ azimuth. Fig. 6(a) shows the diffraction pattern from a clean reconstructed Si(111)-(7×7) surface. The surface was then exposed to air. Prior to Ag deposition the substrate temperature was raised and kept at 550°C. Ag deposition was carried out on the hot substrate. We have varied the thickness of the Ag layer keeping the same experimental condition. But after a certain thickness, the RHEED patterns show that the crystallinity of the Ag-deposited surface is independent of the thickness of the deposited material. The results remain the same for 5 ML, 7 ML and 10 ML of Ag deposition. Fig. 6(b) shows the RHEED pattern from the air-oxidized surface kept at 550°C substrate temperature before deposition. After 2 ML of Ag deposition the diffraction pattern [Fig. 6(c)] remains very similar to that from the oxidized surface. However, some spots from the deposited Ag film are also evident in Fig. 6(c). After deposition of 5 ML of Ag the RHEED pattern essentially comes from the deposited Ag film [Fig. 6(d)]. Fig. 6(e) and (f) show the RHEED patterns after deposition of 7 and 10 ML of Ag respectively. After 5 ML of Ag deposition, the 3D spot feature becomes prominent and this feature does not change with increase in the thickness of the Ag film. After a certain thickness, the pattern does not depend on the thickness of the grown layer. The crystalline quality of the Ag film is observed to be dependent only on the substrate temperature during the deposition. Comparing the RHEED patterns in Fig. 4 and Fig. 6, we notice that for Ag deposition directly on a hot substrate (550°C), well-oriented Ag islands grow directly on the substrate instead of going through a polycrystalline phase.

We analyze the orientation of the Ag nanoislands with respect to the Si substrate via simulation of RHEED patterns from Ag islands in figures 4-6, which are actually transmission electron diffraction patterns as discussed earlier. We generate the simulated patterns based on ref. [35]. These are shown in Fig. 7. Epitaxial growth of Ag on Si(111) surfaces show significant twinning. This means that two types of grains, or twin structures, exist in the film. These two types are conventionally called type A and type B [36]. Both the twins follow the growth: Ag(111) ∥ Si(111); however they differ in the order of stacking. For the A type, the epitaxial relationship is: Ag(111) ∥ Si(111), $[1\,1\,0]_{Ag}$ ∥ $[1\,1\,0]_{Si}$ (i.e., in the plane of the interface: $[1\,\bar{1}\,0]_{Ag}$ ∥ $[1\,\bar{1}\,0]_{Si}$) and for the B type it is Ag(111) ∥ Si(111), $[1\,1\,0]_{Ag}$ ∥ $[1\,1\,4]_{Si}$ (i.e., in the plane of the interface: $[1\,\bar{1}\,0]_{Ag}$ ∥ $[\bar{1}\,1\,0]_{Si}$). This means that in the twinned crystal the Ag unit cell is 180° rotated about the $[1\,1\,1]_{Si}$ axis. We simulate the diffraction patterns for these two types of Ag grains (in our case Ag nanoislands). The comparison of the simulated patterns with the experimental RHEED pattern clearly indicates the epitaxial orientation of the Ag islands. Fig. 7(a) and (b) show the simulated diffraction patterns for the A-type and the B-type twins, respectively with the incident beam along the $[1\,\bar{1}\,0]_{Ag}$ and the $[\bar{1}\,1\,0]_{Ag}$ directions. Fig. 7(c) shows the patterns in (a) and (b) superimposed on each other. The pattern in Fig. 7(c) matches very well with the experimental RHEED pattern [Fig. 7(d)] obtained with the incident beam along $[1\,\bar{1}\,0]_{Si}$. That the incident electron beam is along the $[1\,\bar{1}\,0]_{Ag}$ is additionally corroborated by the value of L/M ratio (1.16) in the diffraction pattern [37] for fcc crystals. This establishes the epitaxial orientation of the Ag islands with the twin structures mentioned above. As further confirmation we discuss the RHEED pattern for the $[1\,1\,\bar{2}]_{Si}$ azimuth along with simulations. The simulated diffraction patterns for both the twin structures mentioned above, with the electron beam along $[1\,1\,\bar{2}]_{Ag}$ and $[\bar{1}\,\bar{1}\,2]_{Ag}$, are identical as seen in Fig. 7(e) and (f). That is, the diffraction from one kind of twin reinforces that from the other. The superimposed pattern of (e) and (f) are shown in Fig. 7(g) with filled and open circles. The observed diffraction pattern [Fig. 7(h)] matches well with the superimposed simulated pattern [Fig. 7(g)]. Additionally, the values of M/N (= 1.63) and L/N (= 1.92) confirm that the beam is incident along $[1\,1\,\bar{2}]_{Ag}$ or its equivalent directions as expected for an fcc crystal [37]. Therefore, the RHEED patterns establish the epitaxial orientation of the Ag nanoislands with the relationship: Ag(111) ∥ Si(111), $[1\,\bar{1}\,0]_{Ag}$ ∥ $[1\,\bar{1}\,0]_{Si}$ ($[1\,1\,\bar{2}]_{Ag}$ ∥ $[1\,1\,\bar{2}]_{Si}$) and Ag(111) ∥ Si(111), $[\bar{1}\,1\,0]_{Ag}$ ∥ $[1\,\bar{1}\,0]_{Si}$ ($[\bar{1}\,\bar{1}\,2]_{Ag}$ ∥ $[1\,1\,\bar{2}]_{Si}$). This is equivalent to the orientations mentioned above for the A- and B-type twinned epitaxial structures. From Fig. 7(d) we notice that the intensities of diffraction spots from each twin are comparable, indicating their comparable probability of existence in the Ag nanoislands. Almost comparable number of A-type and B-type grain growth was also observed for Ag growth on Si(111) surfaces [36,38]. The epitaxial orientation of the Ag islands with the A-type structure is schematically shown in Fig. 8. The weak spots in the RHEED patterns are indeed very weak and can be associated with a very small number of Ag(100) islands as observed in ref. [36,38] for Ag growth on Si(111) surfaces.

Because of the minimum surface energy, the most preferred orientation of Ag islands is Ag(111). However, on an oxidized Si surface, along with the preferential Ag(111) orientation, islands with random azimuthal orientations are expected to grow. It is quite interesting to note that the twinned Ag nanocrystals, as observed by RHEED, have grown only with preferential azimuthal orientations, very similar to the case of epitaxial growth of Ag on Si(111) surfaces. It is not very difficult to understand this. For the growth of an epitaxial layer, long range order on the substrate surface is necessary. For the present case of nanoisland growth, the contact area of the islands with the substrate surface is very small with the lateral dimension of a few nanometers (see Fig. 9), which is about the size of 2-3 (7×7) unit cells (one side of the (7×7) unit cell is 2.7 nm). Apparently the short range (7×7) order on the oxide surface is enough to influence the epitaxial orientation of the Ag islands grown on it.

Several cases of epitaxial growth on oxidized Si surfaces have been reported. Ge epitaxial islands have been grown on a pre-grown ultrathin $SiO_x$ layer on Si(111)-(7×7) surfaces at an elevated temperature (550°C) [39-41]. However, in this case, voids are formed in the oxide layer and Ge epitaxial islands grow within the voids in contact with the Si surface. Epitaxial growth of Ag on native-oxide-covered Si substrates has been reported for radio-frequency magnetron sputtering growth, where energetic bombardment of the substrate by sputtered Ag particles helps remove the native oxide at deposition temperatures $\geq$ 200°C; at 550°C or above the native oxide is completely removed [42,43]. Removal of the native oxide apparently allows the growth of epitaxial Ag islands in contact with the underlying Si. In our case, the Ag islands grow as in epitaxial growth, however, they are not in direct contact with the Si substrate; the oxide layer is intact between the Ag islands and Si, as revealed in the XTEM images presented latter.

As STM looks at the projection of the islands on the surface, very often, this technique can neither provide the correct shape of the islands nor can give the correct base diameter. Also, it can not provide the information about the environment at the interface between the Ag islands and the $SiO_x$/Si layer. In our XTEM studies [Fig. 9], where the cross-sectional view from the side of the islands is observed, we notice that the Ag islands are resting on the oxide layer and they are nearly spherical. The islands have grown only on the $SiO_x$ surface and they have no contact with the underlying Si(111) surface. It is important to obtain this information about the interface region. Details about the island-substrate contact region, such as, whether the islands have contact with Si or not, affect the electronic energy levels in quantum confinement of electrons in nanoislands [40,41].

From the XTEM micrographs in Fig. 9(a) and (b), we notice that the lateral density of islands in Fig. 9(a) is much larger than that in Fig. 9(b). As STM and XTEM have probed smaller regions of the sample, in order to reveal if there is a large lateral nonuniformity of island density at larger length scales, we have taken SEM images from the sample grown at 550°C. These images are shown in Fig. 10, which reveal more or less uniform lateral distribution of islands. This also indicates the uniformity of the oxide growth obtained by air exposure.

## 4. Summary and Conclusions

Air-oxidized Si(111)-(7×7) surfaces have been studied by RHEED and STM. Analysis of Fourier transforms of the STM images has shown the presence of the 1/7$^{th}$ spots indicating the presence of a short range (7×7) order on the oxide surface. Some fractional order spots have also been seen in the RHEED pattern from the air-oxidized surface. Growth of Ag on these air-oxidized surfaces has been investigated. On an oxidized surface, which has a lower surface free energy compared with a clean Si surface, island growth of the deposited Ag is expected. Deposition of Ag at RT on the air-oxidized surfaces has shown Debye-Scherrer rings in the RHEED pattern indicating the presence of randomly oriented Ag islands. Growth at 550°C and for coverages > 2 ML has shown an orientation of Ag islands similar to the case of epitaxial growth of Ag on Si(111) surfaces – a predominant orientation $[1\bar{1}0]_{Ag} \parallel [1\bar{1}0]_{Si}$, $(111)_{Ag} \parallel (111)_{Si}$ along with a twin $[\bar{1}10]_{Ag} \parallel [1\bar{1}0]_{Si}$, $(111)_{Ag} \parallel (111)_{Si}$. The presence of such preferred orientations in the Ag islands in spite of the presence of a 2-3 nm oxide layer between Ag and Si is unexpected. The cross-sectional TEM images clearly show that there is no direct contact between the Ag islands and Si and the intervening oxide layer is intact. One may wonder if the short range (7×7) order present on the oxide layer influences the orientation of the Ag islands. In any case, for the growth of islands of nanoscale size, a long range order on the substrate surface is not necessary for the orientation of the islands to be influenced. The dimension of the contact region of the islands with the oxide surface is ≤ 10 nm or the size of about four (7×7) unit cells. Growth of Ag islands with epitaxial orientation was also observed on air-oxidized Si(100)-(2×1) surfaces. The role of short range order on substrates for the self-assembled growth of nanostructures needs to be investigated more extensively.

**Figure captions**

Fig. 1. STM images of a clean Si(111) surface (a) with monatomic steps and terraces [Bias voltage, $V_s$ = 2.2 V, tunneling current, I = 0.19 nA] and (b) showing (7×7) surface reconstruction [Bias voltage, $V_s$ = 2.2 V, tunneling current, I = 0.18 nA]. Scan areas are (a) 500×500 nm$^2$ and (b) 20×20 nm$^2$. (c) Surface after air-exposure showing that surface steps are preserved [Bias voltage, $V_s$ = 2.2 V, tunneling current, I = 0.19 nA]. (d) After deposition of 10 ML Ag at RT on this surface showing densely packed Ag islands [Bias voltage, $V_s$ = 2.2 V, tunneling current, I = 0.18 nA].

Fig. 2. STM images of the air-exposed Si(111)-(7×7) surfaces of different surface scan areas and their corresponding Fourier transformed images. Scan areas are (a) 300×300 nm$^2$ and (b) 150×150 nm$^2$. The 1/7$^{th}$ fractional spots corresponding to the periodic (7×7) reconstructed surface are observed. The 1/7$^{th}$ spots in the FT of the image in (b) are shown by arrows. Compare the spots with those in Fig. 3.

Fig. 3. STM images of the clean Si(111)-(7×7) reconstructed surface and their corresponding Fourier transformed images. Scan areas are (a) 300×300 nm$^2$ and (b) 150×150 nm$^2$. Positions of the 1/7$^{th}$ fractional spots corresponding to the periodic (7×7) reconstructed surface are marked by double-circle in the FT of (a) and single-circle in the FT of (b). Some higher order spots are also marked by single-circle in the FT of (a). Up to 4/7$^{th}$ order spots are seen in the FT of the image in (b).

Fig. 4. RHEED images from a surface after 10 ML of Ag deposition on air-oxidized Si(111)-(7×7) surfaces at different substrate temperatures. The incident beam is parallel to the [1 1 $\bar{2}$] orientation of Si. RHEED pattern from (a) a clean Si(111)-(7×7) surface, (b) the same surface oxidized upon air-exposure, (c-f) the oxidized surface following Ag deposition [(c) at RT, (d) at 350°C, (e) at 400°C and (f) at 550°C].

Fig. 5. RHEED images from a surface after 10 ML of Ag deposition on air-oxidized Si(111)-(7×7) surfaces at different substrate temperatures. The incident beam is parallel to the [1 $\bar{1}$ 0] orientation of Si. RHEED pattern from (a) a clean Si(111)-(7×7) surface, (b) the same surface

oxidized upon air-exposure, (c-f) the oxidized surface following Ag deposition [(c) at RT, (d) at 350°C, (e) at 400°C and (f) at 550°C].

Fig. 6. RHEED images from a surface with increasing thickness of Ag layer at a substrate temperature of 550°C. The incident beam is parallel to the [1 1 $\bar{2}$] orientation of Si. RHEED pattern from (a) a clean Si(111)-(7×7) surface, (b) the same surface when oxidized upon air-exposure and kept at 550°C substrate temperature before deposition, (c-f) the oxidized surface following Ag deposition with different coverages [(c) 2 ML, (d) 5 ML, (e) 7 ML and (f) 10 ML].

Fig. 7. Simulated diffraction pattern for the [1 $\bar{1}$ 0] azimuth from an fcc Ag crystallite: (a) A-type (b) B-type (twin), (c) superimposed diffraction pattern from coexistent A- and B-type structures. (d) Experimental diffraction pattern. The value of L/M (= 1.16) is as expected for an incident beam along [1 $\bar{1}$ 0]$_{Ag}$ or its equivalent directions. Simulated diffraction patterns for the [1 1 $\bar{2}$] azimuth from an fcc Ag crystallite: (e) A-type, (f) B-type, (g) superimposed diffraction pattern from coexistent A- and B-type structures. (h) Experimental diffraction pattern. The values of M/N (= 1.63) and L/N (= 1.92) are as expected for an incident beam along [1 1 $\bar{2}$] or its equivalent directions for an fcc crystal.

Fig. 8. The epitaxial crystallographic orientation of Ag nanoislands with respect to those of the Si(111) substrate is schematically shown for the A-type islands: [1 1 1]$_{Ag}$ ∥ [1 1 1]$_{Si}$, [1 $\bar{1}$ 0]$_{Ag}$ ∥ [1 $\bar{1}$ 0]$_{Si}$, [1 1 $\bar{2}$]$_{Ag}$ ∥ [1 1 $\bar{2}$]$_{Si}$. For the twinned or B-type structure (not shown): [1 1 1]$_{Ag}$ ∥ [1 1 1]$_{Si}$, [$\bar{1}$ 1 0]$_{Ag}$ ∥ [1 $\bar{1}$ 0]$_{Si}$, [$\bar{1}$ $\bar{1}$ 2]$_{Ag}$ ∥ [1 1 $\bar{2}$]$_{Si}$.

Fig. 9. XTEM micrographs obtained from a 10 ML Ag-deposited film on an air-oxidized Si(111)-(7×7) surface at 550°C growth temperature.

Fig. 10. SEM images from a 10 ML Ag film deposited at 550°C on an air-exposed Si(111)-(7×7) surface.

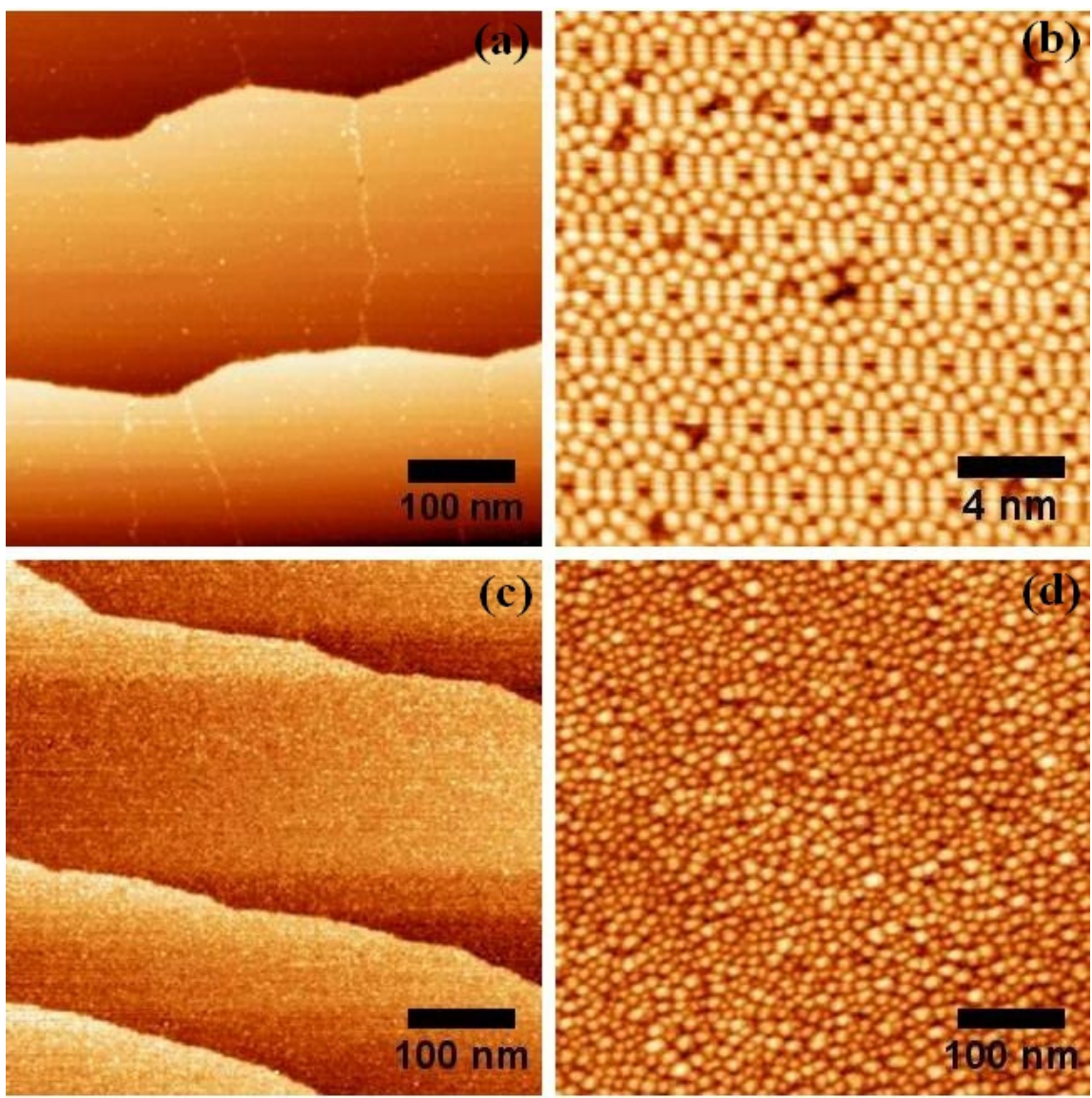

**Fig. 1.** STM images of a clean Si(111) surface (a) with monatomic steps and terraces [Bias voltage, $V_s$ = 2.2 V, tunneling current, I = 0.19 nA] and (b) showing (7×7) surface reconstruction [Bias voltage, $V_s$ = 2.2 V, tunneling current, I = 0.18 nA]. Scan areas are (a) 500×500 $nm^2$ and (b) 20×20 $nm^2$. (c) Surface after air-exposure showing that surface steps are preserved [Bias voltage, $V_s$ = 2.2 V, tunneling current, I = 0.19 nA]. (d) After deposition of 10 ML Ag at RT on this surface showing densely packed Ag islands [Bias voltage, $V_s$ = 2.2 V, tunneling current, I = 0.18 nA].

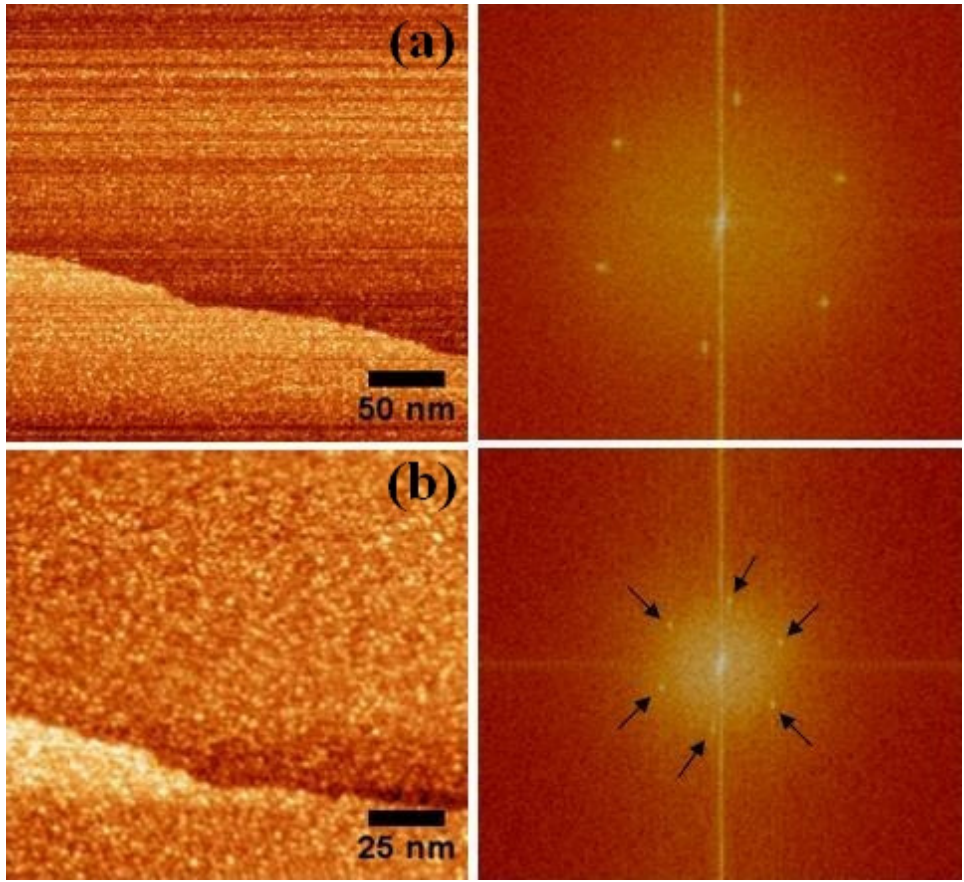

**Fig. 2.** STM images of the air-exposed Si(111)-(7×7) surfaces of different surface scan areas and their corresponding Fourier transformed images. Scan areas are (a) 300×300 nm$^2$ and (b) 150×150 nm$^2$. The 1/7$^{th}$ fractional spots corresponding to the periodic (7×7) reconstructed surface are observed. The 1/7$^{th}$ spots in the FT of the image in (b) are shown by arrows. Compare the spots with those in Fig. 3.

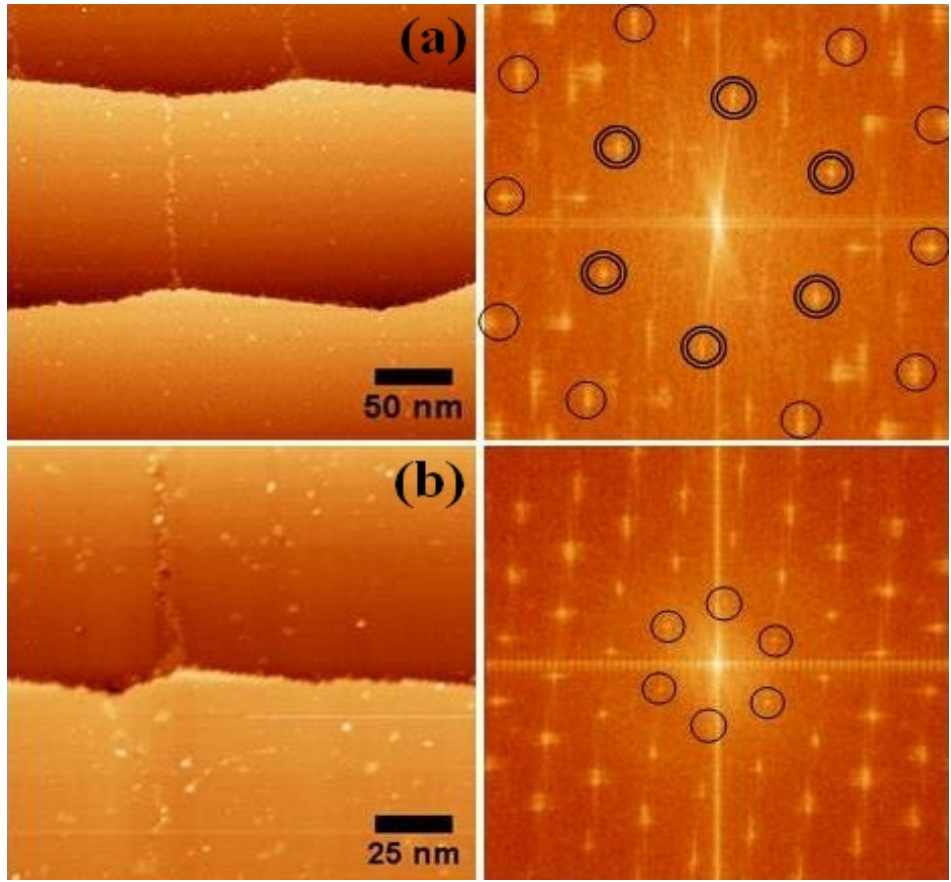

**Fig. 3.** STM images of the clean Si(111)-(7×7) reconstructed surface and their corresponding Fourier transformed images. Scan areas are (a) 300×300 nm$^2$ and (b) 150×150 nm$^2$. Positions of the 1/7$^{th}$ fractional spots corresponding to the periodic (7×7) reconstructed surface are marked by double-circle in the FT of (a) and single-circle in the FT of (b). Some higher order spots are also marked by single-circle in the FT of (a). Up to 4/7$^{th}$ order spots are seen in the FT of the image in (b).

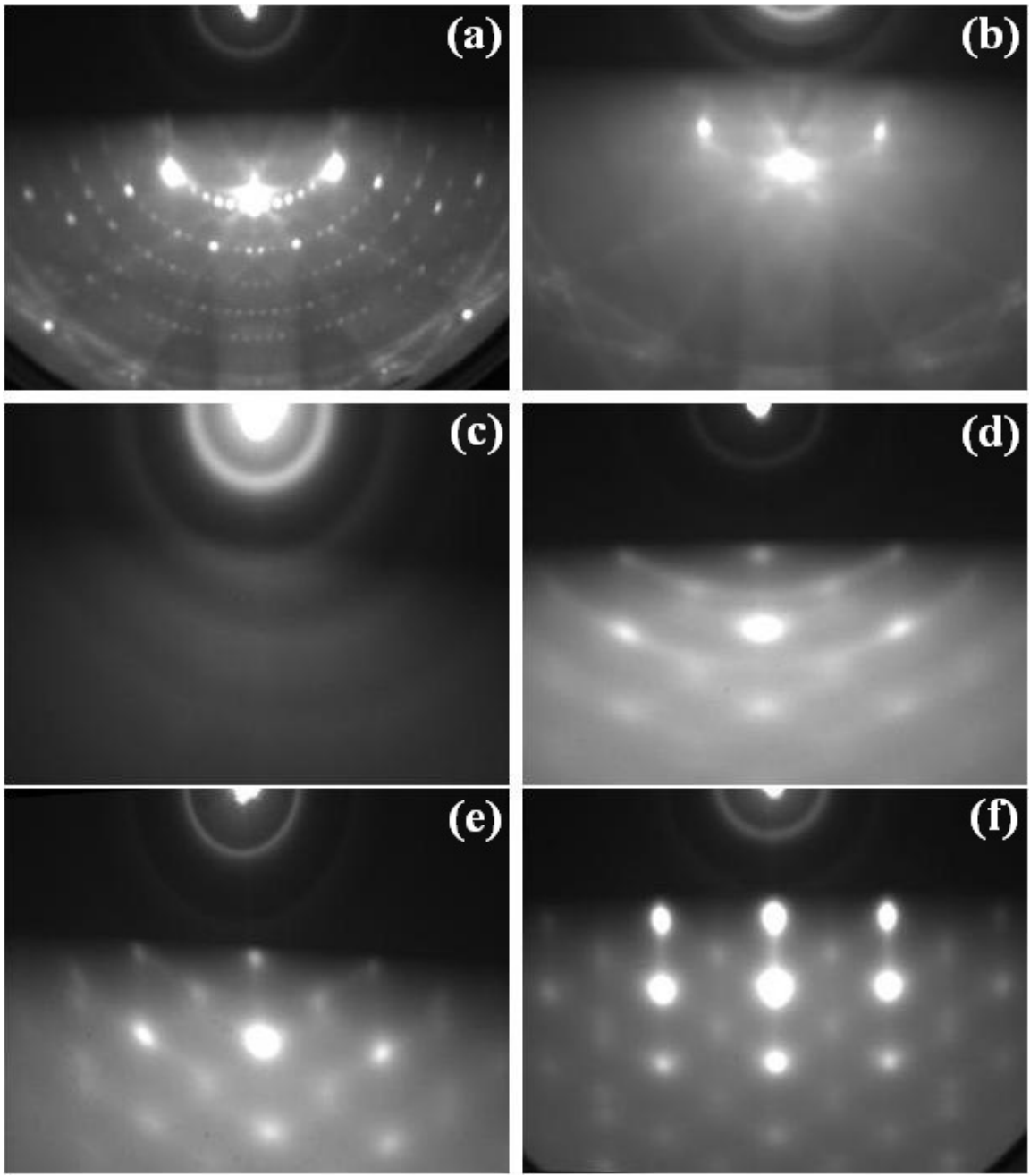

**Fig. 4.** RHEED images from a surface after 10 ML of Ag deposition on air-oxidized Si(111)-(7×7) surfaces at different substrate temperatures. The incident beam is parallel to the [1 1 $\bar{2}$] orientation of Si. RHEED pattern from (a) a clean Si(111)-(7×7) surface, (b) the same surface oxidized upon air-exposure, (c-f) the oxidized surface following Ag deposition [(c) at RT, (d) at 350°C, (e) at 400°C and (f) at 550°C].

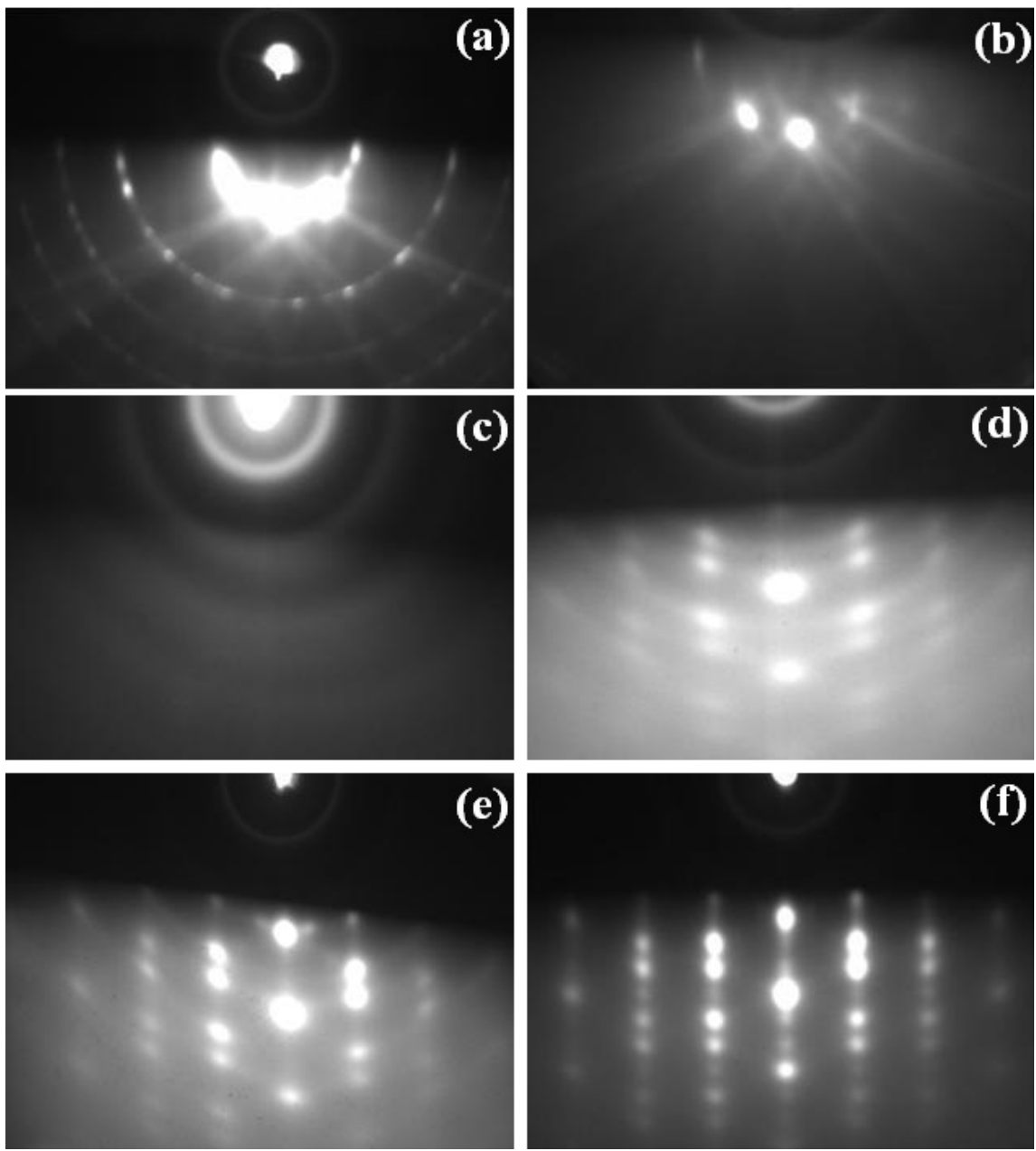

**Fig. 5.** RHEED images from a surface after 10 ML of Ag deposition on air-oxidized Si(111)-(7×7) surfaces at different substrate temperatures. The incident beam is parallel to the [1 $\bar{1}$ 0] orientation of Si. RHEED pattern from (a) a clean Si(111)-(7×7) surface, (b) the same surface oxidized upon air-exposure, (c-f) the oxidized surface following Ag deposition [(c) at RT, (d) at 350°C, (e) at 400°C and (f) at 550°C].

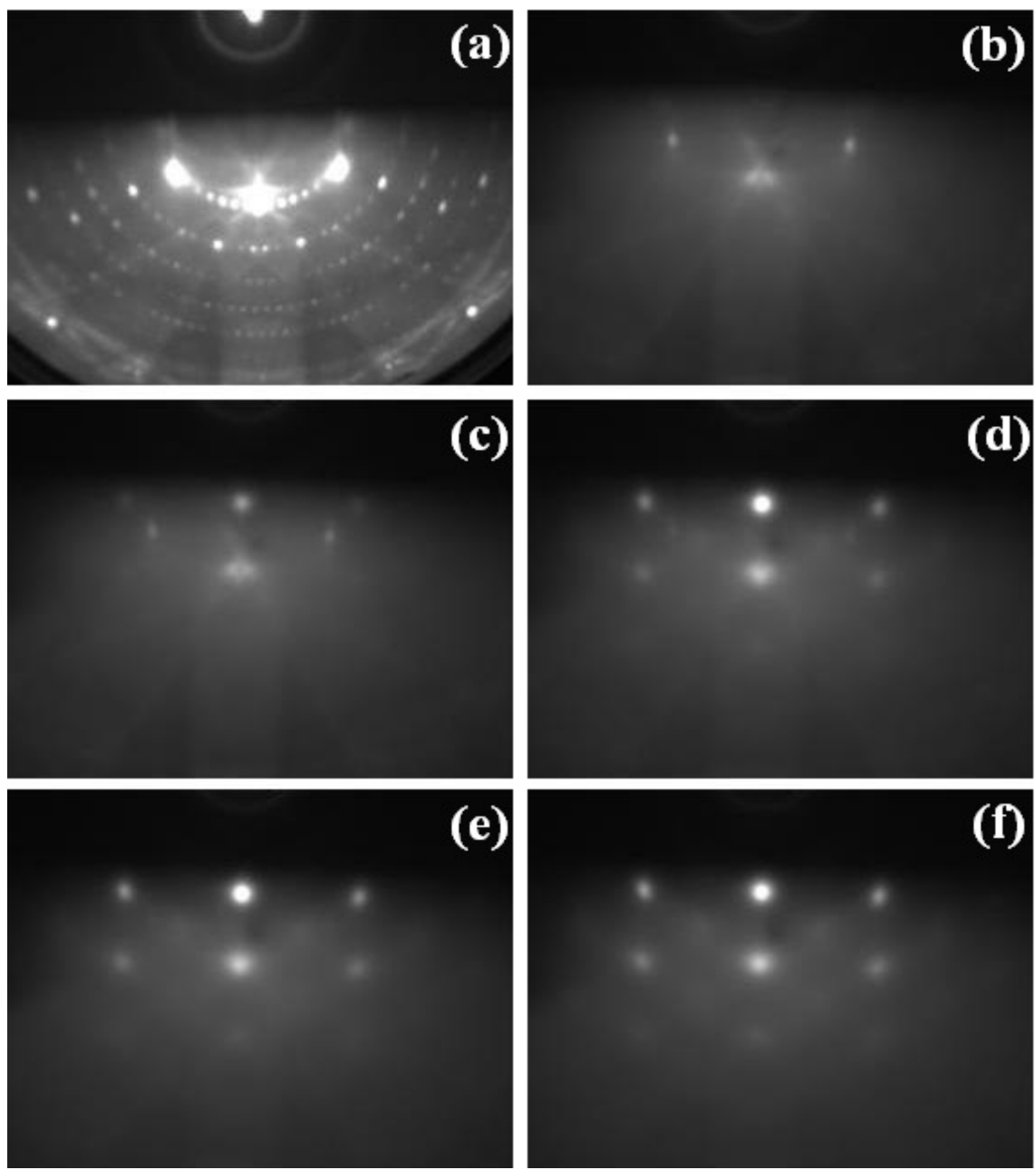

**Fig. 6.** RHEED images from a surface with increasing thickness of Ag layer at a substrate temperature of 550°C. The incident beam is parallel to the [1 1 $\bar{2}$] orientation of Si. RHEED pattern from (a) a clean Si(111)-(7×7) surface, (b) the same surface when oxidized upon air-exposure and kept at 550°C substrate temperature before deposition, (c-f) the oxidized surface following Ag deposition with different coverages [(c) 2 ML, (d) 5 ML, (e) 7 ML and (f) 10 ML].

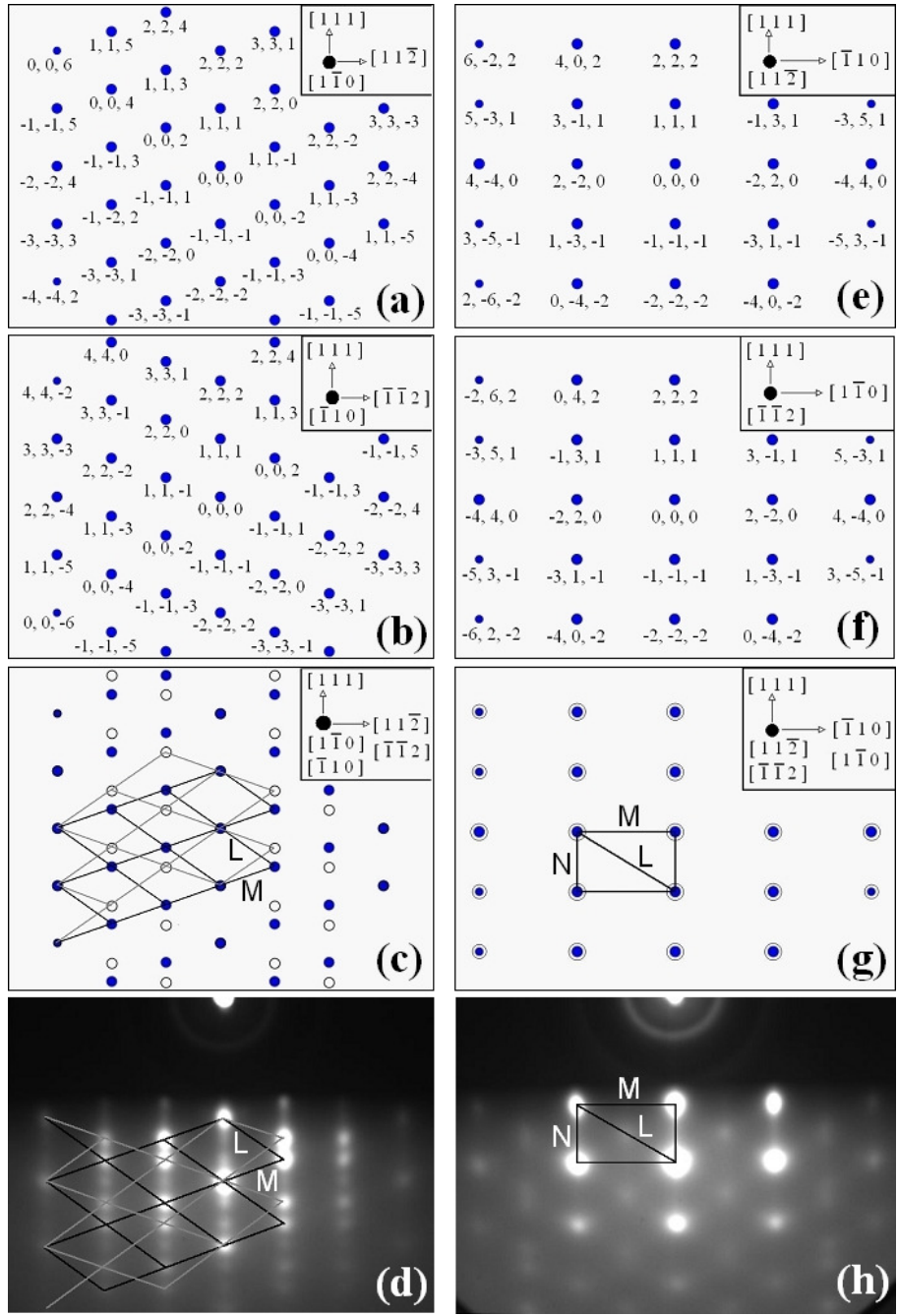

**Fig. 7.** Simulated diffraction pattern for the [1 $\bar{1}$ 0] azimuth from an fcc Ag crystallite: (a) A-type (b) B-type (twin), (c) superimposed diffraction pattern from coexistent A- and B-type structures. (d) Experimental diffraction pattern. The value of L/M (= 1.16) is as expected for an incident beam along [1 $\bar{1}$ 0]$_{Ag}$ or its equivalent directions. Simulated diffraction patterns for the [1 1 $\bar{2}$] azimuth from an fcc Ag crystallite: (e) A-type, (f) B-type, (g) superimposed diffraction pattern from coexistent A- and B-type structures. (h) Experimental diffraction pattern. The values of M/N (= 1.63) and L/N (= 1.92) are as expected for an incident beam along [1 1 $\bar{2}$] or its equivalent directions for an fcc crystal.

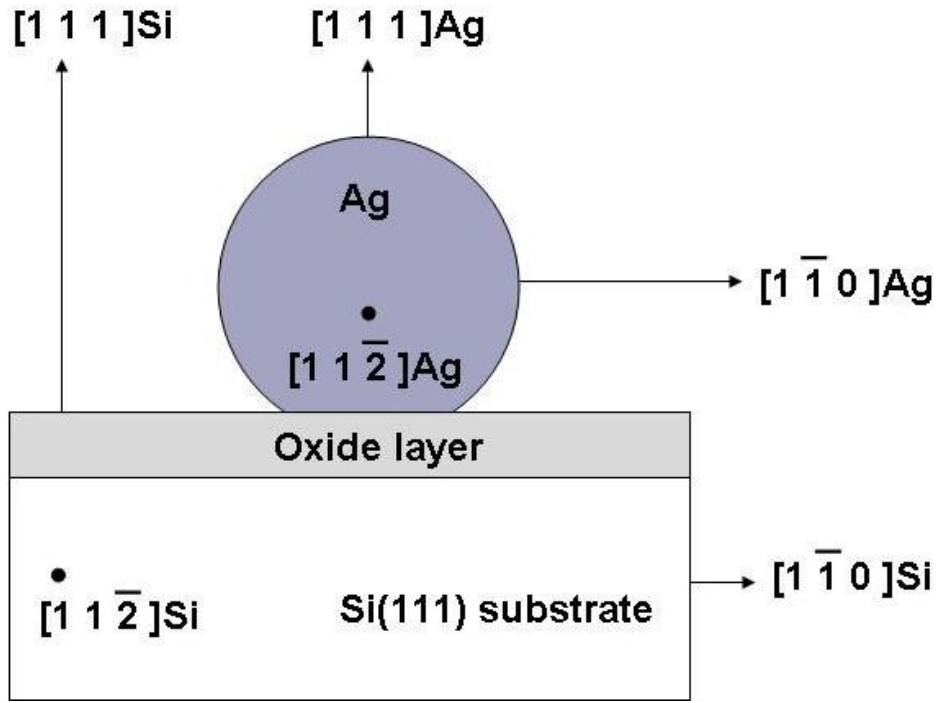

**Fig. 8.** The epitaxial crystallographic orientation of Ag nanoislands with respect to those of the Si(111) substrate is schematically shown for the A-type islands: $[1\ 1\ 1]_{Ag} \parallel [1\ 1\ 1]_{Si}$, $[1\ \bar{1}\ 0]_{Ag} \parallel [1\ \bar{1}\ 0]_{Si}$, $[1\ 1\ \bar{2}]_{Ag} \parallel [1\ 1\ \bar{2}]_{Si}$. For the twinned or B-type structure (not shown): $[1\ 1\ 1]_{Ag} \parallel [1\ 1\ 1]_{Si}$, $[\bar{1}\ 1\ 0]_{Ag} \parallel [1\ \bar{1}\ 0]_{Si}$, $[\bar{1}\ \bar{1}\ 2]_{Ag} \parallel [1\ 1\ \bar{2}]_{Si}$.

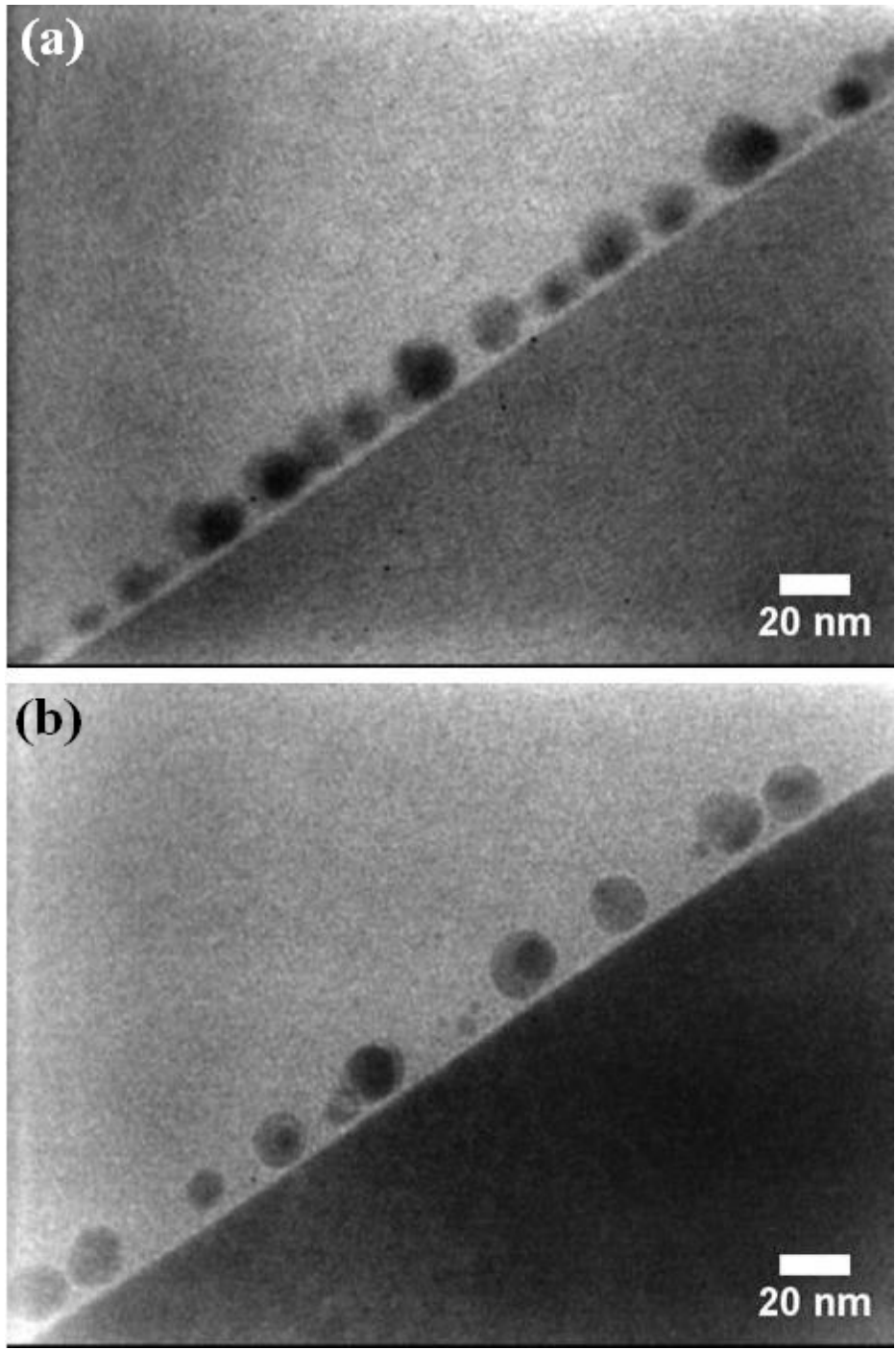

**Fig. 9.** XTEM micrographs obtained from a 10 ML Ag-deposited film on an air-oxidized Si(111)-(7×7) surface at 550°C growth temperature.

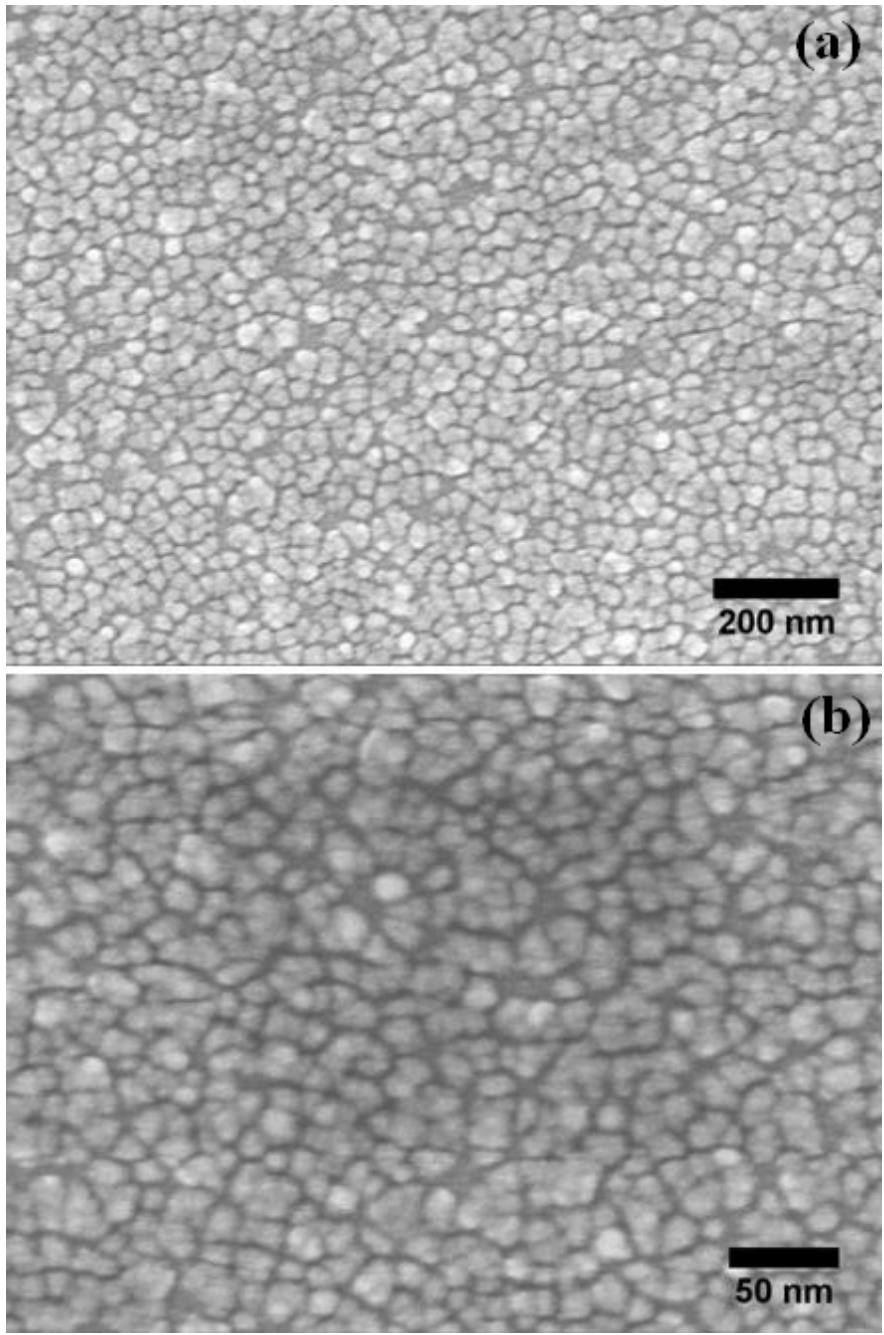

**Fig. 10.** SEM images from a 10 ML Ag film deposited at 550°C on an air-exposed Si(111)-(7×7) surface.